\documentclass[sigconf,anonymous=false,10pt]{acmart}
\usepackage[linesnumbered,ruled,vlined]{algorithm2e}
\usepackage{fvextra}
\usepackage[frozencache=true,cachedir=minted-cache]{minted} 
\usepackage{cuted}
\usepackage{graphicx} 
\usepackage{listings}
\usepackage{hyperref}
\usepackage{enumitem}
\usepackage{booktabs}
\usepackage{multirow}
\usepackage{subfigure} 
\usepackage{todonotes}
\usepackage{enumitem}
\usetikzlibrary{arrows.meta, positioning, calc}

\AtBeginDocument{%
  }

\setcopyright{acmlicensed}
\copyrightyear{2018}
\acmYear{2018}
\acmDOI{XXXXXXX.XXXXXXX}
\acmConference[Conference acronym 'XX]{Make sure to enter the correct
  conference title from your rights confirmation emai}{June 03--05,
  2018}{Woodstock, NY}
\acmISBN{978-1-4503-XXXX-X/18/06}

\begin{document}

\title{DiOMP-Offloading: Toward Portable Distributed Heterogeneous OpenMP}

\author{Baodi Shan}
\email{baodi.shan@stonybrook.edu}
\affiliation{%
  \institution{Stony Brook University}
  \streetaddress{100 Nicolls Road}
  \city{Stony Brook}
  \state{New York}
  \country{USA}
}

\author{Mauricio Araya-Polo}
\affiliation{%
  \institution{TotalEnergies EP Research \& Technology US}
  \streetaddress{P.O. Box 1212}
  \city{Houston}
  \state{Texas}
  \country{USA}
}

\author{Barbara Chapman}
\email{barbara.chapman@stonybrook.edu}
\affiliation{%
  \institution{Stony Brook University}
  \streetaddress{100 Nicolls Road}
  \city{Stony Brook}
  \state{New York}
  \country{USA}
}

\renewcommand{\shortauthors}{Shan et al.}

\begin{abstract}
The two main current trends in HPC are the increasing number of cores -their heterogeneity- and the higher memory bandwidth. The former directly impacts programmability, portability and scalability, and it is the main concern addressed in this work. 
As heterogeneous supercomputing becomes mainstream, traditional hybrid models such as MPI+OpenMP struggle to efficiently manage distributed GPU memory and deliver portable performance.

This paper introduces distributed OpenMP offload (DiOMP-Offloading), a novel framework that unifies OpenMP target offloading with a Partitioned Global Address Space (PGAS) model. DiOMP is built atop LLVM/OpenMP using GASNet-EX or GPI-2 as the communication layer, DiOMP transparently manages global memory regions and supports both symmetric and asymmetric GPU memory allocations. It relies upon OMPCCL, a novel portable layer for collective communication that interfaces seamlessly with vendor-specific libraries. 
Compared to MPI+X approaches, DiOMP achieves superior scalability and programmability for most test cases 
by abstracting away device memory and communication details. 
Experiments across multiple large-scale platforms including NVIDIA A100 and Grace Hopper, and AMD MI250X demonstrate that DiOMP delivers better performance in micro-benchmarks and real-world applications such as matrix multiplication and Minimod. These results indicate that DiOMP has the potential to be part of a more portable, scalable, and efficient future for heterogeneous computing.

\end{abstract}




\maketitle

\section{Introduction}

As high-performance computing (HPC) platforms evolve toward increasingly heterogeneous and large-scale architectures, modern systems are often composed of multi-node, multi-CPU, multi-GPU configurations interconnected by high-speed fabric and networks. In domains such as numerical simulation and artificial intelligence, HPC applications impose demanding requirements on programming models. These include support for high concurrency, complex data distribution across heterogeneous memory hierarchies, and the ability to balance portability, programmability, and performance. While traditional hybrid programming models like MPI+OpenMP have been widely adopted, the emergence of GPUs as primary computational engines in many leadership-class systems—such as OLCF's Frontier and NERSC's Perlmutter—poses significant challenges for developers. These include manual device memory management, explicit control of host-device transfers, and the need to orchestrate high-performance collective operations through vendor-specific communication libraries such as NVIDIA Collective Communication Library (NCCL) or AMD ROCm Communication Collectives Library (RCCL). Such complexities significantly increase development and optimization effort, reduce productivity, and hinder the exploitation and maintainability of hardware capabilities.

To address these challenges, researchers have increasingly explored higher-level programming abstractions that eliminate the need for explicit data placement and communication orchestration. The Partitioned Global Address Space (PGAS) model offers a logical global memory abstraction that simplifies remote memory access in distributed systems. Recent efforts have focused on incorporating GPU memory into this global address space to unify memory access across heterogeneous nodes. However, existing approaches typically depend on specialized APIs (e.g., NVSHMEM) or lack deep integration with directive-based parallel models like OpenMP—models that are critical for incremental code modernization and rapid prototyping.

This paper introduces \textbf{DiOMP-Offloading}, an extension of the DiOMP framework~\cite{diomp}, which unifies PGAS-style data distribution with OpenMP target offloading and enables the integration of GPU device memory into a globally addressable space. All this built atop the LLVM/OpenMP runtime and leveraging GASNet-EX\cite{gasnet} or GPI-2\cite{gpi2} as the communication substrate, DiOMP-Offloading constructs a distributed runtime environment that allows transparent allocation and access to data across multiple GPUs, without requiring explicit device memory management by the user. The runtime uniformly supports both symmetric and non-symmetric heap regions, providing a consistent and efficient access model across node boundaries.

More importantly, the DiOMP-Offloading runtime introduces a \textbf{unified framework} for managing communication and computation across heterogeneous resources. It handles device memory registration, lifecycle, and synchronization in a centralized and coordinated manner, eliminating inconsistencies between memory management and communication semantics found in traditional models, while maximizing resource efficiency.

To support collective communication on GPU-resident data, we further introduce \textbf{OMPCCL} (OpenMP Collective Communication Layer)—a unified abstraction layer that bridges OpenMP programming with vendor-specific communication libraries such as NCCL and RCCL. OMPCCL encapsulates common device-side collective operations (e.g., broadcast, reduce, all-reduce) and provides a clean, portable interface for OpenMP applications to leverage high-performance communication backends. While low-level optimizations—such as topology-aware path selection and zero-copy mechanisms—are handled by the underlying vendor-specific collective libraries, OMPCCL’s contribution lies in enabling these capabilities to be accessed through a standard-compliant and OpenMP-compatible interface for the first time. This design simplify the integration of collective communication in OpenMP offloading workflows and serves as a prototype for future standardization of device-side collective operations within the OpenMP specification. 

DiOMP-Offloading eliminates the dependency on vendor-specific APIs and enables OpenMP to operate transparently in distributed \textbf{CPU+GPU heterogeneous systems}. Unlike traditional MPI-based models that assign one GPU per process—thereby limiting intra-node CPU parallelism—or single-process multi-GPU models that suffer from inefficient communication, DiOMP-Offloading introduces a scalable single-process, multi-GPU initialization strategy. This approach allows OpenMP to fully utilize all CPU threads for host-side computation while maintaining efficient collective communication among GPUs via OMPCCL, even in a multi-GPU single-process setting.

The core contributions of this paper are as follows:

\begin{itemize}[wide]

\item \textbf{Unified runtime for communication and computation:} Device memory registration, allocation, synchronization, and lifecycle management are handled uniformly by the DiOMP-Offloading runtime, supporting efficient coordination between communication and computation.

\item \textbf{Heterogeneous device memory integration}: DiOMP-Offloading supports transparent remote access to both symmetric and asymmetric GPU memory regions, abstracting away differences in allocation origin or scope.

\item \textbf{PGAS integration with OpenMP offloading:} We provide the first unified model that bridges PGAS-style global data distribution with OpenMP target offloading, enabling a topology-aware global address space across heterogeneous nodes.

\item \textbf{Portable device-side collectives via OMPCCL:} We introduce a general-purpose abstraction layer that exposes high-performance collective operations on GPU memory through a portable and OpenMP-compatible API, leveraging vendor-optimized XCCL implementations underneath.

\item \textbf{Efficient multi-GPU OpenMP deployment model:} We present a scalable single-process multi-GPU initialization strategy that retains full OpenMP threading flexibility while preserving communication performance through OMPCCL.
\end{itemize}


The rest of this paper is organized as follows. Section~\ref{sec:background_related} reviews the background on OpenMP, the PGAS model, and the DiOMP runtime. Section~\ref{sec:design} describes the architecture and implementation of DiOMP-Offloading, including its runtime workflow, memory management, and communication model. Section~\ref{sec:exp} presents a detailed evaluation of DiOMP-Offloading using both micro-benchmarks and full applications. Section~\ref{sec:conclusion} concludes the paper and discusses future research directions.
\section{Background and Related Work}
\label{sec:background_related}

\subsection{OpenMP and PGAS Models}

OpenMP~\cite{openmp5} is a widely-used programming model for enabling shared-memory parallelism in high-performance computing. It offers a user-friendly and flexible interface, allowing developers to utilize the parallelism of multi-core processors and shared memory systems. Since the introduction of the \texttt{task} construct in OpenMP 3.0, developers can express independent units of work for concurrent execution, making it particularly suitable for irregular parallelism, recursive algorithms, and applications with complex dependencies. OpenMP 4.0 extended this paradigm by introducing task dependencies and device offloading~\cite{shileitask,arxivbaodi}, enabling code execution on accelerators without requiring vendor-specific APIs~\cite{9741290,9820621}.

In parallel, the Partitioned Global Address Space (PGAS) model~\cite{pgasintro, jungblut2021portable} contrasts with the Message Passing Interface (MPI) by providing a globally accessible memory space partitioned across distributed processing units. PGAS supports one-sided communication operations, such as \textbf{get} and \textbf{put}, allowing asynchronous remote memory access without active target participation. Notable implementations of PGAS include OpenSHMEM, Legion, UPC++, DASH, Chapel, and OpenUH Co-Array Fortran, with GASNet being a widely adopted communication framework~\cite{upcxx,chapel,legion,openshmem}.

Despite these advances, OpenMP and PGAS programming models face challenges in integrating support for heterogeneous systems. For example, OpenSHMEM's lack of accelerator support has limited its applicability in heterogeneous computing environments, despite the introduction of preliminary proposals such as Symmetric Partitions and Memory Spaces~\cite{openshmem}.

\textbf{DiOMP}~\cite{diomp} is a Partitioned Global Address Space (PGAS) model built on top of LLVM/OpenMP and GASNet-EX. By incorporating Remote Memory Access (RMA) operations such as \texttt{put} and \texttt{get}, DiOMP simplifies data movement and synchronization in distributed environments, offering improved programmability compared to traditional MPI+OpenMP approaches. However, the current implementation~\cite{diomp} does not support heterogeneous architectures, such as GPUs. This paper presents DiOMP-Offloading, an extension of DiOMP re-designed to enable efficient offloading of OpenMP programs in distributed heterogeneous systems.

\subsection{Challenges and Advances in Distributed OpenMP and PGAS Integration}

Research into extending OpenMP and PGAS models for distributed and heterogeneous architectures has gained momentum. Remote OpenMP offloading frameworks~\cite{remote,optremote,mpiremote} and cluster-based solutions such as OMPC~\cite{OMPC} explore scalable alternatives to MPI+OpenMP, but often face bottlenecks due to centralized task scheduling. Similarly, SHMEM-based frameworks, including NVIDIA's NVSHMEM~\cite{nvshmem2023}, provide GPU-specific solutions but lack portability and flexibility for broader heterogeneous systems.

Hybrid approaches combining PGAS models and accelerator support, such as OpenSHMEM+OpenMP Target Offloading, have been explored but face significant limitations. Lu et al.~\cite{10024665} rely on an outdated LLVM Offload version and treat OpenMP and OpenSHMEM as separate models, which restricts their ability to efficiently handle heterogeneous computations. Similarly, NVIDIA’s NVSHMEM~\cite{nvshmem2023}, while effective for NVIDIA GPUs, lacks generality and necessitates manual CUDA management, thereby limiting its applicability in broader heterogeneous systems. The work of ~\cite{Nakao19} focuses on the combination of PGAS and OpenMP target offloading within a single node
\section{Design of DiOMP-Offloading}
\label{sec:design}
In this section, we introduce the design and implementation of DiOMP-Offloading, a runtime system tailored for scalable and efficient execution of OpenMP programs in distributed heterogeneous environments. Building upon the foundation of DiOMP, DiOMP-Offloading extends the original design with GPU offloading capabilities and deeply integrated communication semantics. We begin by presenting the system’s overall workflow, highlighting its unified approach to memory management and communication. Then, we elaborate on the key architectural components—including the global memory abstraction, hierarchical P2P data movement, and OMPCCL-enabled collective communication—which together enable DiOMP-Offloading to deliver high performance and programmability across large-scale GPU clusters.
\subsection{Workflow of DiOMP Offloading}
\label{sec:workflow}

DiOMP Offloading is built upon the LLVM infrastructure by extending the OpenMP target offloading implementation (\texttt{libomptarget}) and integrating it with high-performance communication middleware such as GASNet-EX and GPI-2. This work constructs a unified and comprehensive framework that enables efficient GPU-accelerated computation and high-throughput inter-node data movement.

\begin{figure*}[bhtp]
    \vspace{-8 mm}
    \centering
    \subfigure[Workflow of the OpenMP Target + MPI]{
        \includegraphics[width=.48\textwidth]{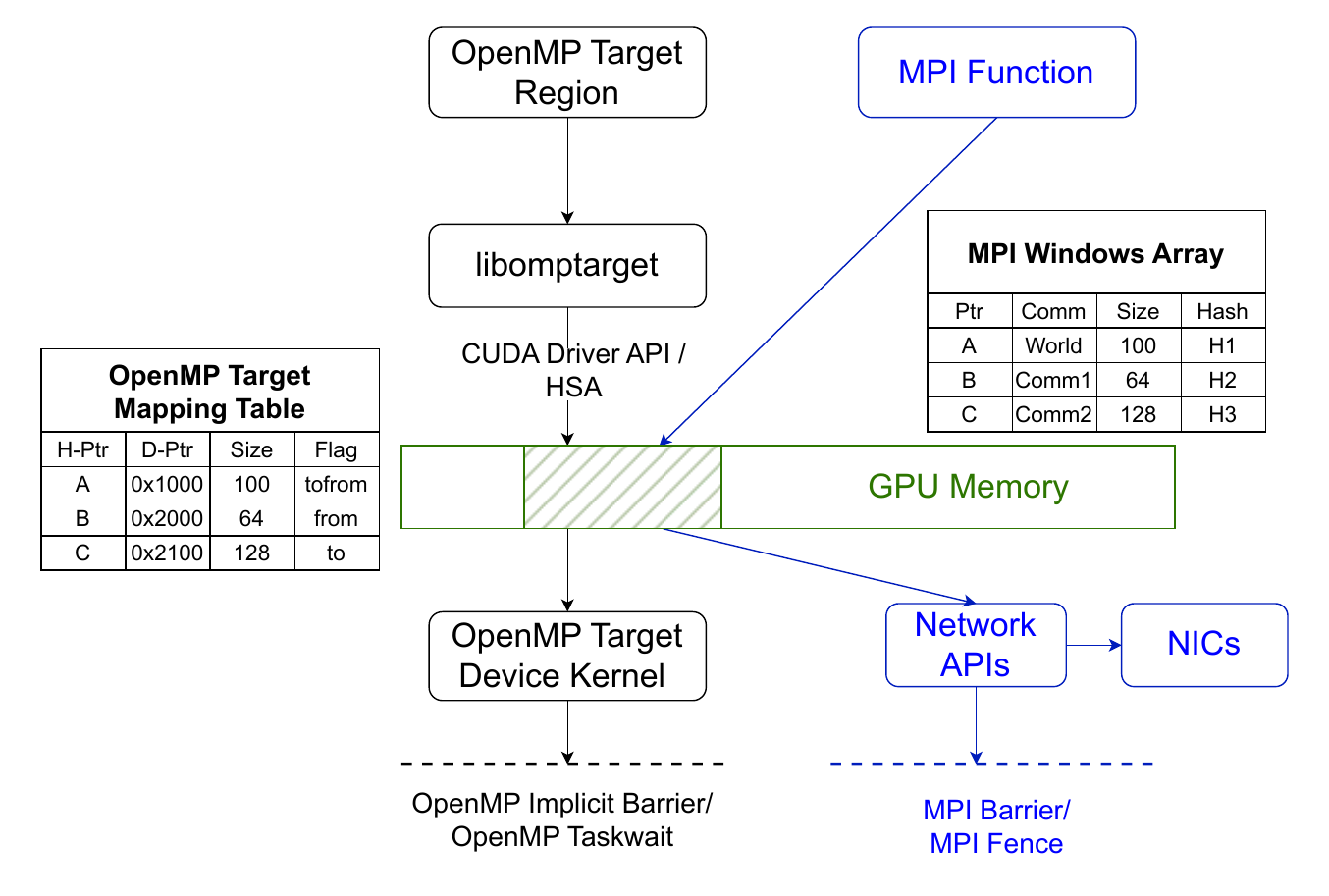}
    }
    \subfigure[Workflow of DiOMP]{
        \includegraphics[width=.48\textwidth]{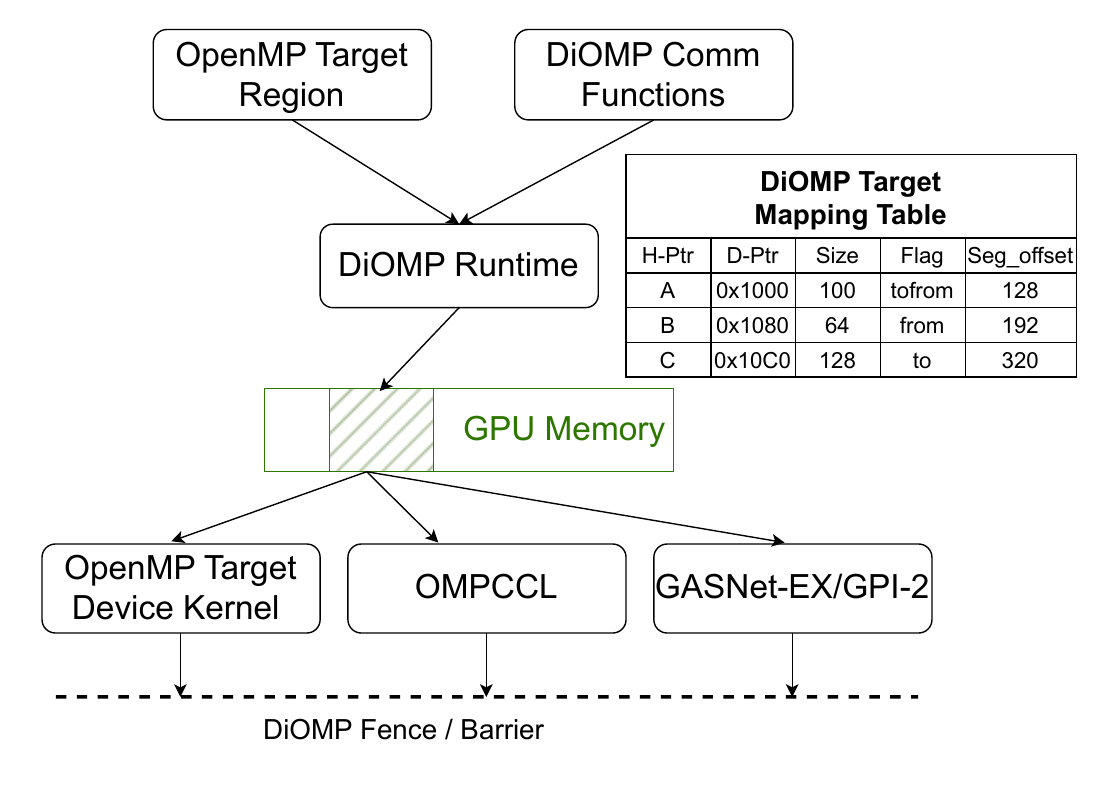}
    }
    \vspace{-4 mm}
    \caption{Comparison of data management and communication workflows between OpenMP Target + MPI and DiOMP-Offloading. \textbf{(a)} In the OpenMP Target + MPI approach, libomptarget and MPI manage GPU memory separately, each maintaining its own metadata and performing independent memory registration via distinct APIs (e.g., CUDA Driver and MPI windows). This separation leads to redundant memory handling, inconsistent synchronization (e.g., OpenMP implicit barrier vs. MPI fence), and uncoordinated data lifecycles. \textbf{(b)} DiOMP-Offloading provides a unified runtime that integrates OpenMP target regions and communication functions. It manages a centralized mapping table and coordinates memory registration and synchronization, avoiding duplication and ensuring consistency across layers.}
    \label{fig:workflow}
\end{figure*}

\autoref{fig:workflow} illustrates the key differences between DiOMP’s memory management mechanism and the traditional MPI +
\texttt{libomptarget} architecture. In \texttt{libomptarget}, device memory allocation relies on the underlying CUDA Driver API or HSA Runtime implementation, and each \texttt{target} region independently manages its data mapping and memory lifecycle. Even under communication models that support CUDA-aware MPI or PGAS by default, users must explicitly register device memory into MPI windows or the PGAS global space. This leads to redundant memory management, duplicated mappings, and potential consistency issues across different system modules.
In contrast, the DiOMP runtime takes over the device memory allocation process and constructs memory regions through a unified memory allocation interface. These regions are directly allocated in the global segment managed by GASNet-EX (or GPI-2), using strategies such as a linear heap allocator or a buddy allocator to build a unified PGAS global space with cross-node accessibility.
More importantly, the allocated memory is jointly accessible and managed by the \texttt{libomptarget}, the point-to-point (P2P) communication path of GASNet-EX (or GPI-2), and the collective communication components of OMPCCL. This enables zero-copy sharing of data and co-management of memory lifecycles between communication and offloading.
This design achieves a deep integration of memory management and communication semantics by tightly coupling memory management, communication mechanisms, and computation scheduling at the system architecture level, enabling them to share metadata, resource states, and execution contexts. 

Ultimately, DiOMP-Offloading builds a unified execution model tailored for heterogeneous systems, enabling integrated scheduling of communication, computation, and memory resources, thereby significantly improving system communication efficiency and scalability.

\subsection{Global Memory Management and Hierarchical Data Transfer}

This subsection describes the global memory management strategy of DiOMP and its topology-aware, hierarchical point-to-point communication mechanism.

 \begin{figure}[htbp]
    \centering
    \vspace{-4 mm}
    \includegraphics[width=.48\textwidth]{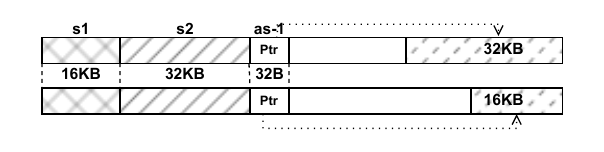}
    \vskip -0.2in
    \caption{Symmetric and asymmetric memory allocation in DiOMP Offloading.}
    \label{fig:memory}
\end{figure}

DiOMP follows the design principles of mainstream PGAS models by allocating symmetric global memory regions across participating nodes, enabling transparent remote data access via \texttt{put} and \texttt{get} operations. On the CPU side, users can allocate memory in the global address space manually using \texttt{omp\_alloc}. On the GPU side, as described in~\autoref{sec:workflow}, the DiOMP runtime intercepts the default memory allocation routines from \textit{libomptarget}, redirecting all OpenMP-mapped device memory allocations into a globally managed device memory segment under DiOMP's control.

In the current design, DiOMP-Offloading adopts a collective memory allocation mechanism within the global address space, where all participating nodes are required to coordinate during the allocation phase. However, unlike traditional PGAS models such as OpenSHMEM---which enforce strict symmetry in global memory allocation (e.g., via \texttt{shmem\_alloc})---DiOMP-Offloading supports both \emph{symmetric} and \emph{asymmetric} allocation modes, providing increased flexibility while preserving communication correctness.

In the \textbf{symmetric allocation} mode, all ranks allocate an identical amount of global memory. This symmetry ensures consistent offset mappings across nodes: a given pointer corresponds to the same relative position (offset) within the global memory segments on all nodes. DiOMP leverages this property to simplify address translation for remote access: the target address on a remote node is computed as the base address of that node’s global memory segment plus the local pointer’s offset. As illustrated in Figure~\ref{fig:memory}, nodes A and B allocate symmetric memory blocks S1 and S2, of size 16KB and 32KB respectively. When node A intends to access node B’s memory region S2, it can compute the corresponding remote address by applying the local offset of S2 (on node A) to the base address of node B’s global memory segment. This mechanism enables a logically coherent global address space and supports efficient one-sided communication.

In contrast, \textbf{asymmetric allocation} allows each rank to allocate differing amounts of global memory. However, such allocations invalidate the consistent offset assumption, rendering the direct offset-based address translation mechanism unusable. To resolve this issue, DiOMP introduces the concept of a \emph{second-level pointer abstraction}, illustrated in Figure~\ref{fig:memory}(as-1). A second-level pointer is essentially a 32-byte pointer wrapper, uniformly allocated across all ranks to preserve global alignment. The actual asymmetric memory is then allocated at the end of the global segment, and the second-level pointer is updated to reference this non-uniform memory region.

This indirection introduces a new communication challenge: since the data address must be dereferenced through a remote pointer, remote memory operations generally require two communication steps. The first fetches the second-level pointer value, and the second performs the actual data transfer. To mitigate the performance penalty of repeated two-stage communication, DiOMP implements a \emph{remote pointer caching mechanism}, which maintains a mapping of previously fetched remote second-level pointers. This cache reduces redundant communication and improves runtime efficiency. Furthermore, since DiOMP manages both memory allocation and deallocation centrally, it ensures that each second-level pointer’s cache entry is valid throughout the lifetime of its corresponding memory allocation.

While DiOMP-Offloading supports both symmetric and asymmetric allocation, from a performance optimization standpoint, symmetric allocation remains the preferred approach under the PGAS model. Therefore, in memory-abundant scenarios, we encourage developers to emulate symmetry through manual padding techniques, thereby retaining the benefits of offset-based address translation and maximizing one-sided communication efficiency.

DiOMP further introduces a \textbf{topology-aware, hierarchical communication framework} for point-to-point data transfers. The runtime dynamically detects GPU topology and selects the optimal communication path accordingly:

\begin{itemize}[wide, topsep=0pt]
    \item For GPUs on different nodes, DiOMP uses GASNet-EX for inter-node communication. We also provide GPI-2 for InfiniBand environment.
    \item For GPUs on the same node but belonging to different processes, DiOMP employs \textbf{Inter-Process Communication (IPC)} mechanisms (e.g., CUDA IPC Memory Handles or HIP IPC) to achieve efficient intra-node transfers.
    \item For GPUs that support GPUDirect P2P, DiOMP invokes \textit{cudaDeviceEnablePeerAccess} or \textit{hipDeviceEnablePeerAccess} to enable direct memory fabric-based transfers, minimizing latency and maximizing bandwidth.
    \item To minimize synchronization latency caused by the mismatch between network and device-side events during remote memory operations, DiOMP synchronizes both GASNet-EX (or GPI-2) and CUDA/HIP stream events in a unified polling loop as part of the synchronization process. This coordination ensures efficient overlap of communication and computation, eliminating unnecessary stalls.
\end{itemize}

Special attention has been given to optimizing \textbf{event and stream management} within each node. DiOMP adopts a unified strategy to minimize overhead and maximize responsiveness by coordinating GPU streams and communication events from GASNet-EX. The following runtime techniques are employed:

\begin{itemize}[wide] 
\item \textbf{Lazy Allocation:} Streams are not preallocated but instantiated on demand to reduce idle resource usage. 
\item \textbf{Stream Reuse:} If idle streams exist in the pool, they are reused instead of creating new ones. 
\item \textbf{Bounded Concurrency:} We introduce a threshold \texttt{MAX\_ACTIVE\_STREAMS} to control the number of active concurrent streams. When the threshold is reached, the runtime performs \emph{partial synchronization}---only half of the completed streams are synchronized and released, while the remaining active streams continue execution. This policy sustains \textbf{pipeline throughput and responsiveness} while minimizing scheduling and memory pressure on the GPU. 
\item \textbf{Hybrid Event Polling:} To address the asynchronous between network events and device-side streams, DiOMP uses a unified polling mechanism during \texttt{ompx\_fence} operations. The runtime simultaneously polls GASNet-EX completion events and CUDA/HIP stream events in a coordinated loop, ensuring timely progress of both communication and computation. This effectively eliminates stalls caused by mismatched event readiness and enhances RMA efficiency. 
\end{itemize}

Through these mechanisms, DiOMP achieves a runtime-level integration of memory management, communication path selection, and resource scheduling. Specifically:

\begin{enumerate}[wide]
    \item \emph{Unified memory view underpins communication structure:} Communication paths operate directly over the PGAS memory space established by the memory management system, eliminating explicit registration or copying.
    \item \emph{Scheduling adapts to memory and communication state:} Stream allocation and reuse decisions depend on memory usage, and communication path choices take resource availability into account.
    \item \emph{Shared metadata and resource context:} Each memory block is associated with a stream, and modules share execution context to avoid redundant scheduling and module-level isolation.
\end{enumerate}

The notion of ``deep integration’’ in DiOMP-Offloading is grounded in a concrete architectural design that unifies data layout, runtime semantics, and hardware resource orchestration. This unified approach enables improved communication efficiency and system scalability across large-scale heterogeneous computing environments.

\subsection{OMPCCL and DiOMP Group}

Modern HPC systems have highly heterogeneous architectures with intra-node multi-accelerator configurations, but achieving efficient collective communication remains a fundamental technical challenge. 
Mainstream MPI implementations—including those with support for GPU-aware communication—have made notable progress in adapting to heterogeneous platforms. Unfortunately, significant limitations persist in terms of performance optimization and architectural portability.
For instance, certain MPI implementations such as MVAPICH-GDR have demonstrated performance advantages over NCCL in specific scenarios~\cite{osu-nccl}. However, their lack of flexibility and limited cross-platform compatibility impose substantial constraints. A representative example can be found in clusters built on the NVIDIA Grace Hopper architecture (see Section~\ref{sec:exp}), where MVAPICH-GDR neither provides precompiled binaries nor makes its source code publicly available. These limitations significantly hinder its applicability and portability on emerging hardware platforms, thereby reducing its practical utility in next-generation HPC environments.

To achieve unified and efficient collective communication in the DiOMP-Offloading framework, we introduce the abstraction of \textit{DiOMP Group}. Conceptually similar to the \texttt{Communicator} in MPI, a DiOMP Group partitions the global communication domain into smaller, logically distinct subgroups. This allows for fine-grained control over collective communication and resource management. Each group is represented by a lightweight handle of type \texttt{ompx\_group\_t}, and can be dynamically created, merged, or split during runtime to adapt to the evolving needs of multi-phase or task-based programs.

For example, synchronization primitives such as \texttt{ompx\_ba-\\rrier()} and \texttt{ompx\_fence()} can be scoped to a specific DiOMP Group by passing an additional \texttt{ompx\_group\_t} parameter. This design avoids unnecessary global synchronization and allows finer control over communication domains. Furthermore, DiOMP supports group recomposition, where multiple existing groups can be dynamically merged into a new logical group at runtime. This enables modular and reusable communication patterns that can be flexibly adapted to different program phases, each with its own granularity and communication topology.

Building on this group abstraction, we implement OMPCCL, a collective communication layer that supports both intra-group and global collective operations. OMPCCL provides a unified, high-level abstraction over vendor-specific collective communication library, which are critical for achieving high-performance communication and topology-aware optimization in GPU-based systems. The design of OMPCCL abstracts the low-level communication details while preserving the efficiency and scalability offered by NCCL/RCCL. The entire setup and management of communication resources are handled transparently by the DiOMP runtime system, enabling consistent usage across different hardware configurations.

During the initialization phase, the runtime automatically establishes collective communication channels, including the generation and coordination of UniqueIDs. These identifiers are broadcast across processes via a CPU-side communication mechanism to ensure global consistency and correctness. This design not only leverages the optimized transport and topology discovery mechanisms of NCCL/RCCL but also enables seamless integration of device-level communication into the DiOMP execution model through the OMPCCL abstraction.

To simplify the use of collective operations in OpenMP, we propose a set of new, extended directives and runtime functions in DiOMP. Specifically, we introduce a custom pragma syntax such as:
\begin{verbatim}
#pragma ompx target device_bcast(var, ompx_group_t)
\end{verbatim}
This pragma allows developers to explicitly specify device-side broadcast operations within an OpenMP \texttt{target} region, where \texttt{ompx\_group\_t} defines the scope of the broadcast as a particular DiOMP Group. Although this syntax is not part of the current OpenMP standard, it follows OpenMP’s directive-based design and has been prototyped in our compiler and runtime extensions. Notably, the OpenMP Language Committee has also expressed interest~\cite{deSupinski2023TR12} in standardizing device-side collective operations in future versions of the specification, making our design a potential reference for upcoming proposals.

In addition, we provide equivalent DiOMP C/C++ APIs such as 
\begin{verbatim}
ompx_bcast(void* ptr, size_t size, ompx_group_t group)
\end{verbatim}, which offer the same functionality through function calls. This dual interface design—based on both pragmas and explicit APIs—ensures compatibility with diverse programming preferences and provides a foundation for potential future standardization of device-level collective operations in OpenMP. 

The integration of OMPCCL into DiOMP not only enables efficient collective communication across GPU devices, but also introduces new opportunities for harnessing the hierarchical organization and architectural heterogeneity inherent in modern supercomputing systems. In particular, DiOMP adopts a hierarchical device binding strategy that allows each rank to be associated with either a single accelerator or a set of accelerators within a node. Binding a single device per rank preserves compatibility with conventional MPI-based models and benefits from established communication optimizations. Conversely, binding multiple devices to a single rank enhances intra-node resource utilization and facilitates more efficient host-side orchestration of heterogeneous workloads. 

In traditional MPI or PGAS systems, collective communication is typically defined at the rank or processing element (PE) level, with no finer granularity. When a single rank manages multiple devices, collective operations such as AllReduce cannot be performed atomically across all devices, which complicates the communication logic and may increase overall latency. A common workaround is to implement a hierarchical AllReduce scheme, where an intra-rank AllReduce is first performed among the devices, followed by an inter-rank AllReduce across nodes. However, such a two-phase approach often interferes with the internal scheduling strategies of modern MPI libraries which already provide highly optimized, topology-aware, GPU-direct, and RDMA-enabled implementations. Manually breaking the collective into multiple stages not only introduces extra synchronization overhead but can also degrade performance, especially under PCIe-based or suboptimal interconnects. While assigning each device to its own MPI rank may resolve communication granularity issues and allow direct use of \texttt{MPI\_Allreduce}, it introduces a new challenge: fragmented CPU control. With multiple ranks controlling separate devices, the CPU-side computation and orchestration capabilities are partitioned and cannot be globally coordinated, limiting the efficiency of host-device collaboration—a problem that becomes particularly pronounced in applications requiring tight CPU-GPU coordination. DiOMP addresses this tradeoff by decoupling communication groups from rank boundaries and enabling collective operations over arbitrary subsets of devices via the OMPCCL layer. Furthermore, OMPCCL leverages the topology-aware initialization mechanisms provided by NCCL and RCCL to automatically detect device interconnects and select optimized transport paths accordingly.

\section{Experimental Evaluation}
\label{sec:exp}
In this section, we first describe the experimental setup and highlight a hardware limitation relevant to our evaluation. We then present microbenchmark results for both point-to-point and collective communication to analyze the performance characteristics of DiOMP-Offloading. Finally, we evaluate DiOMP using two real-world applications to demonstrate its practical effectiveness and scalability in heterogeneous HPC environments.

\subsection{Experimental Setup}

The experimental design in this study encompasses a variety of heterogeneous HPC platforms, aiming to comprehensively evaluate the performance, compatibility, and adaptability of DiOMP across diverse hardware and software environments.

\begin{figure*}[htbp]
    \centering
    \subfigure[Slingshot 11 + A100]{
        \includegraphics[width=.31\textwidth]{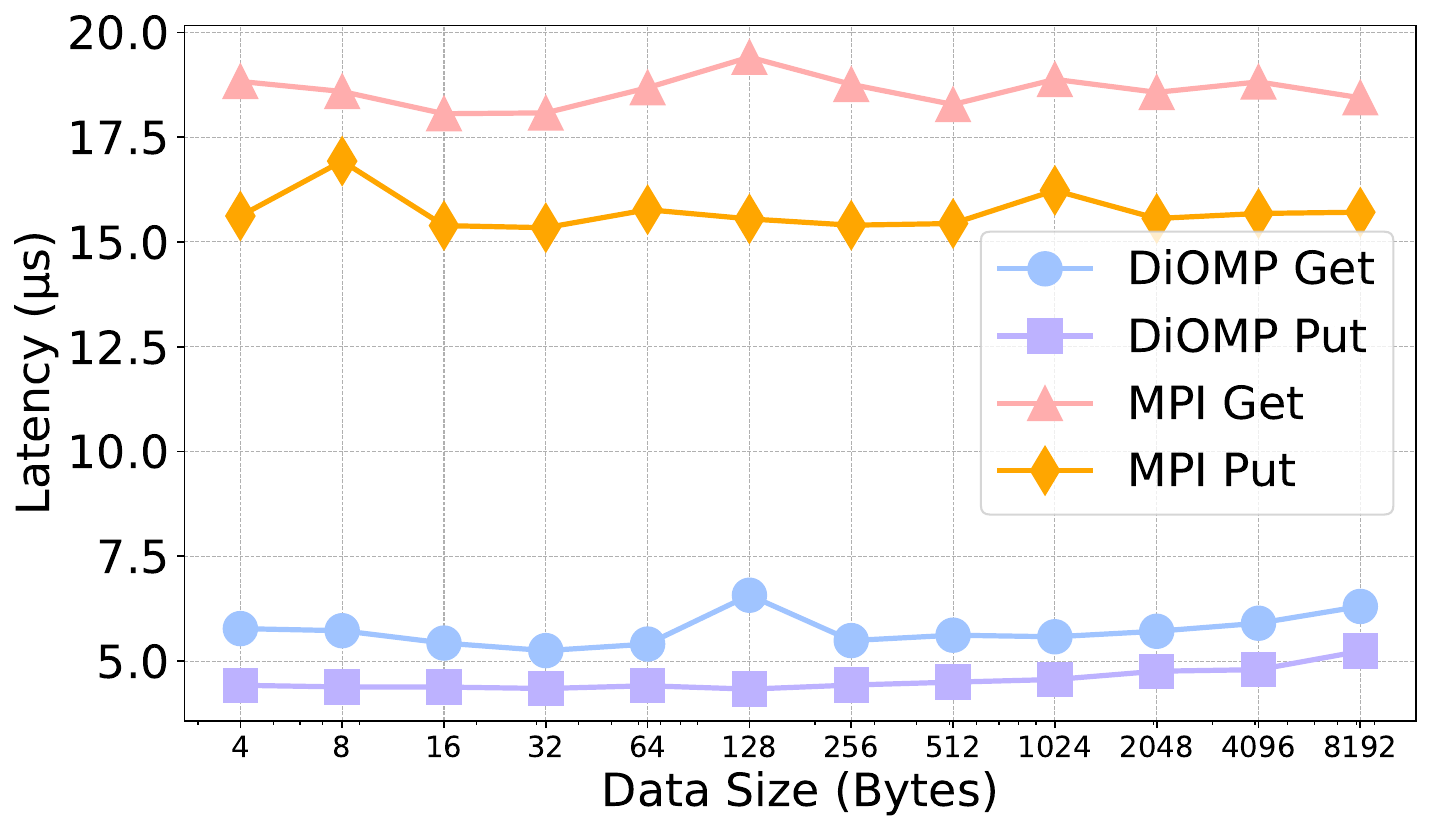}
    }
    \subfigure[Slingshot 11 + MI250X]{
        \includegraphics[width=.31\textwidth]{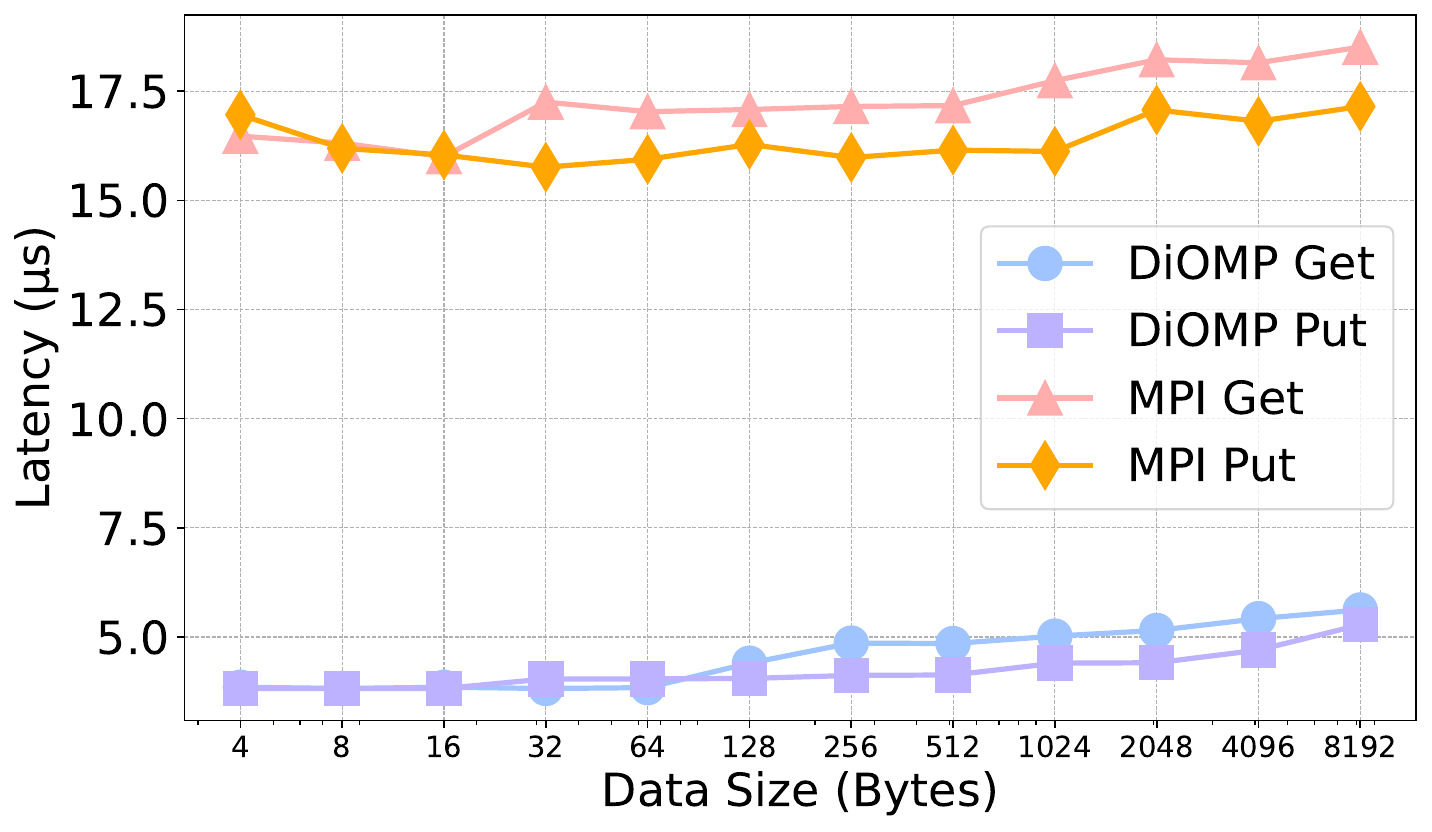}
    }
    \subfigure[NDR InfiniBand + Grace Hopper]{
        \includegraphics[width=.31\textwidth]{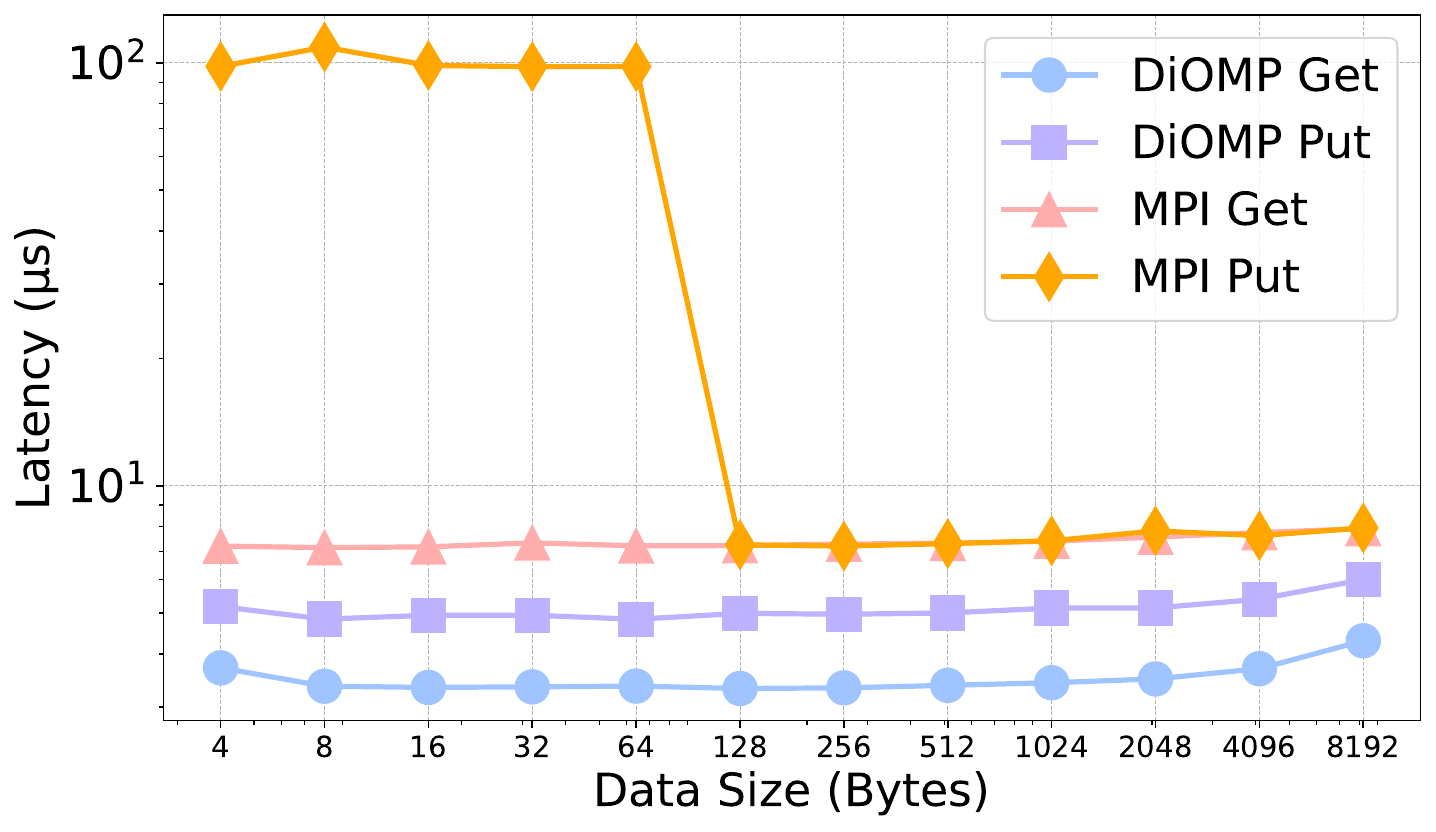}
    }
    \vspace{-5 mm}
    \caption{Latency comparison of DiOMP and MPI operations using InfiniBand and HPE Slingshot 11 from 4 bytes to 8KB. Lower is better.}
    \label{fig:lat}
\end{figure*}

\begin{figure*}[htbp]
    \centering
    \subfigure[Slingshot 11 + A100*]{
        \includegraphics[width=.31\textwidth]{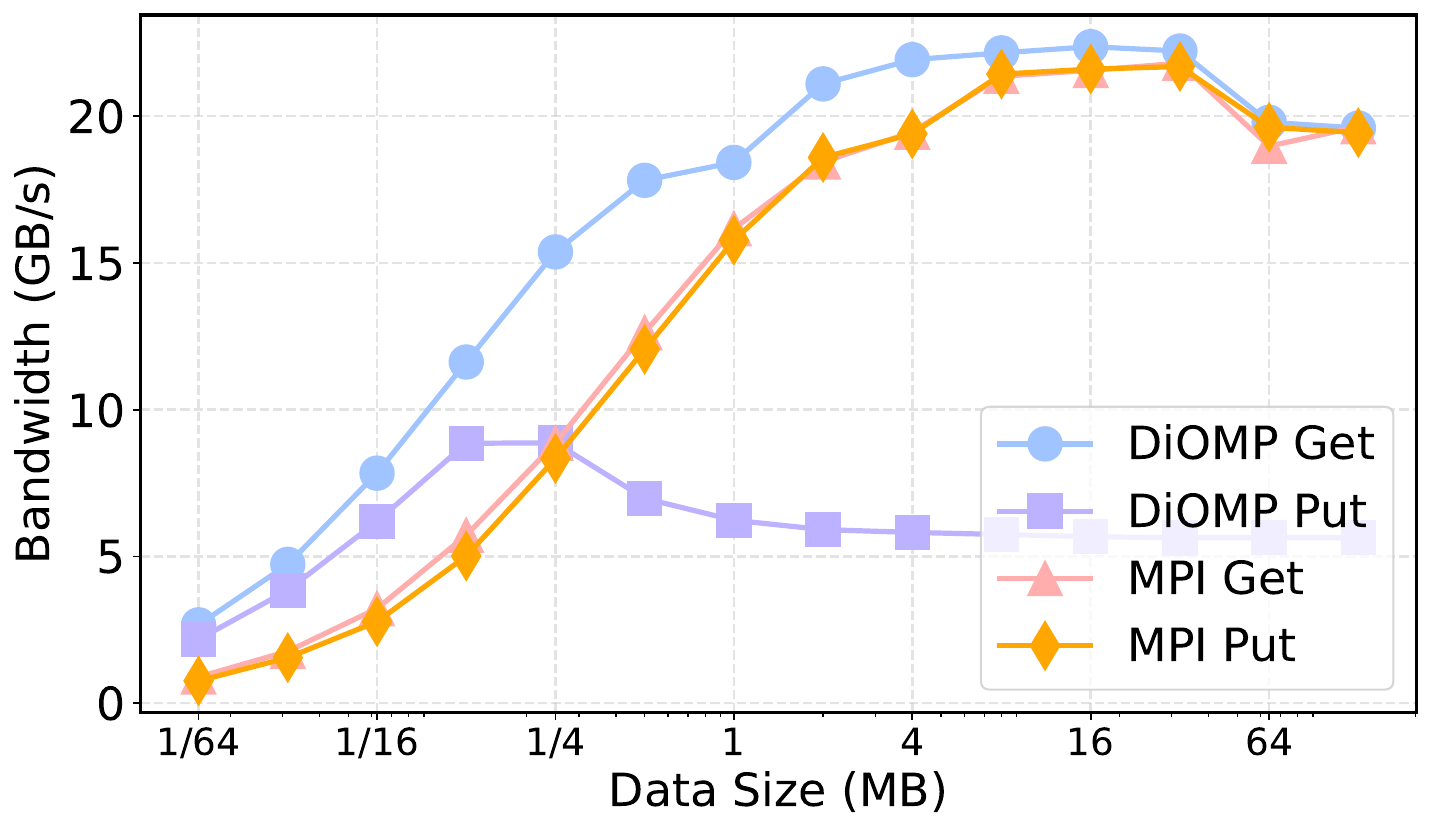}
    }
    \subfigure[Slingshot 11 + MI250X]{
        \includegraphics[width=.31\textwidth]{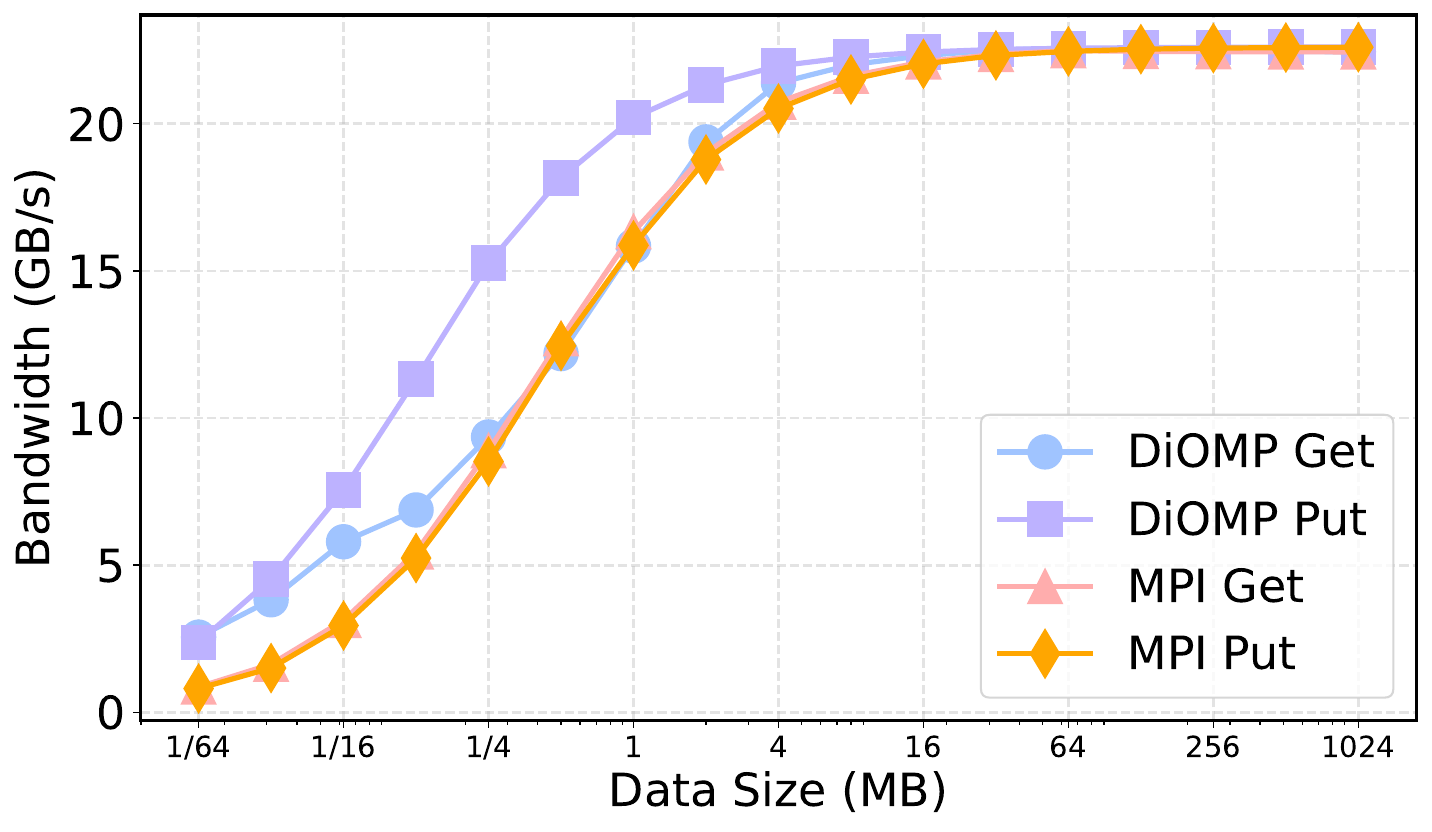}
    }
    \subfigure[NDR InfiniBand + Grace Hopper]{
        \includegraphics[width=.31\textwidth]{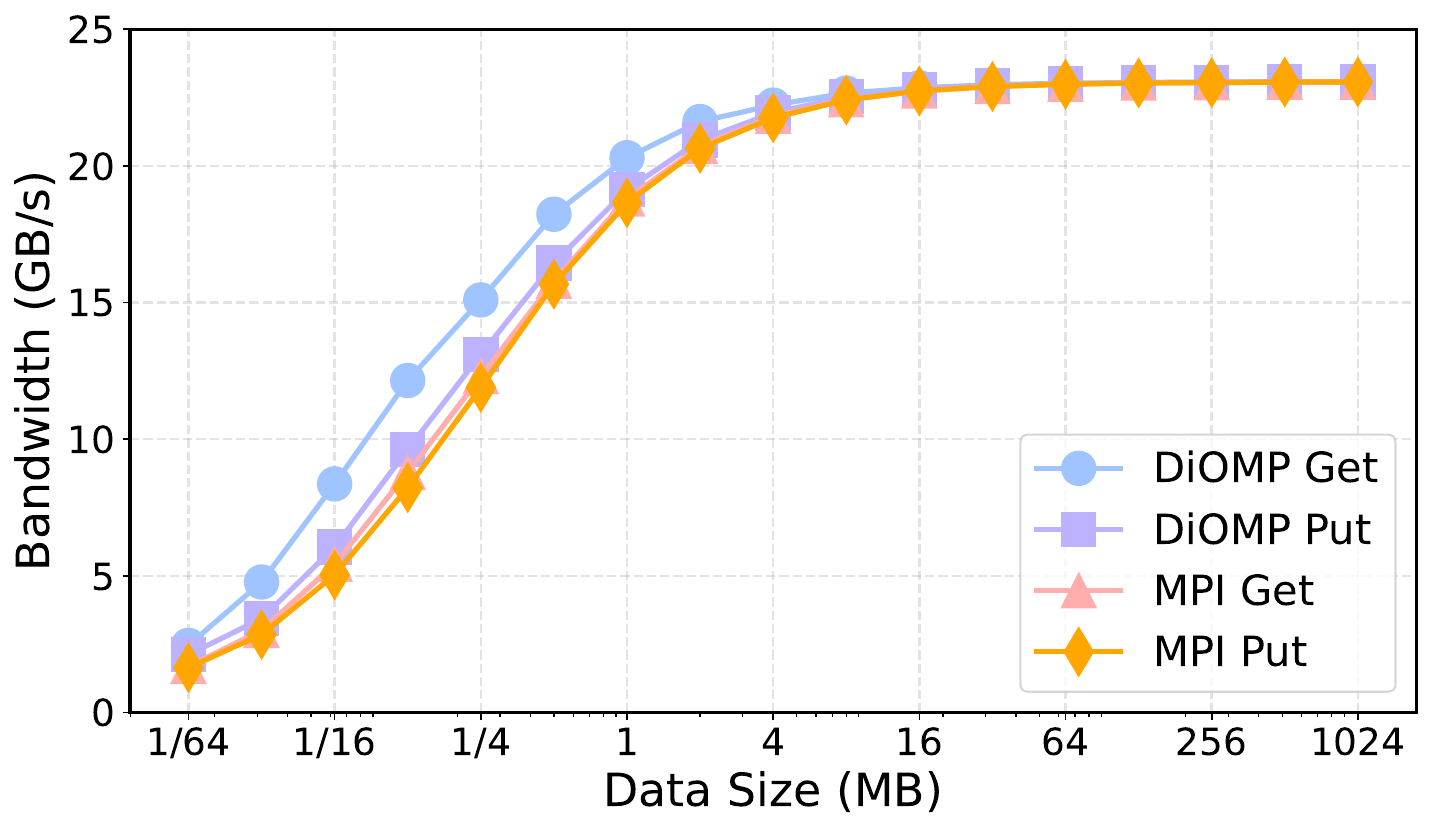}
    }
    \vspace{-5 mm}
    \caption{Bandwidth comparison of DiOMP and MPI operations using InfiniBand and HPE Slingshot 11 across varying data sizes. *The anomalous behavior of DiOMP Put in Slingshot 11 + A100 has been addressed below. Higher is better.}
    \label{fig:bw}
\end{figure*}

\noindent On the hardware side, the experiments were conducted on the following three platforms:

\textbf{Platform A}: A supercomputer consisting of AMD EPYC 7763 CPUs and NVIDIA \textbf{A100} GPUs. Each node includes four \textbf{HPE Slingshot 11} NICs, each providing 200 Gb bandwidth. 

\textbf{Platform B}: A HPC cluster consisting of AMD EPYC 7A53 CPUs and AMD \textbf{MI250X} GPUs. Each node includes four AMD MI250X GPUs and four \textbf{HPE Slingshot 11} NICs each providing 200 Gb bandwidth. Note that a single MI250X itself has two graphics compute dies (GCDs), so one node of Platform C has a total of eight devices for OpenMP purpose. 

\textbf{Platform C}: A HPC cluster based on the NVIDIA Grace Hopper Superchip \textbf{(GH200)}. Each node is equipped with an NVIDIA Grace CPU and Hopper GPU. The nodes are interconnected via a \textbf{200 Gb NDR InfiniBand} network. 

\noindent On the software side:

\textbf{Communication Middleware}: The communication middleware of DiOMP is based on the latest GASNet-EX version 2024.5.0, with network interface adaptations tailored to each platform. Platform A and B employs the OpenFabrics Interfaces (OFI) library, Platform C utilizes the OpenFabrics Verbs (IBV) Network API. 
We also provide an implementation using GPI-2 as the communication middleware; however, it currently supports only InfiniBand environments.

\textbf{Compiler Toolchain}: DiOMP leverages a customized LLVM compiler (based on commit \texttt{f8cc509}) to support specific memory allocation and communication optimizations.

\textbf{Benchmarking Environment}: For comparative experiments, Platform A and B employ HPE Cray MPICH as the baseline, while Platform C uses OpenMPI. These benchmarks provide a basis for evaluating the improvements in DiOMP’s communication performance relative to traditional MPI implementations. Through this combination of hardware and software experimental designs, we ensure a comprehensive assessment of DiOMP, covering multi-node distributed communication performance, single-node high-bandwidth memory computation capabilities, and compatibility with different hardware platforms.

\subsection{Micro-benchmark - Point to Point}
\begin{figure}[htbp]
    \vspace{-2 mm}
    \centering
    \includegraphics[width=.48\textwidth]{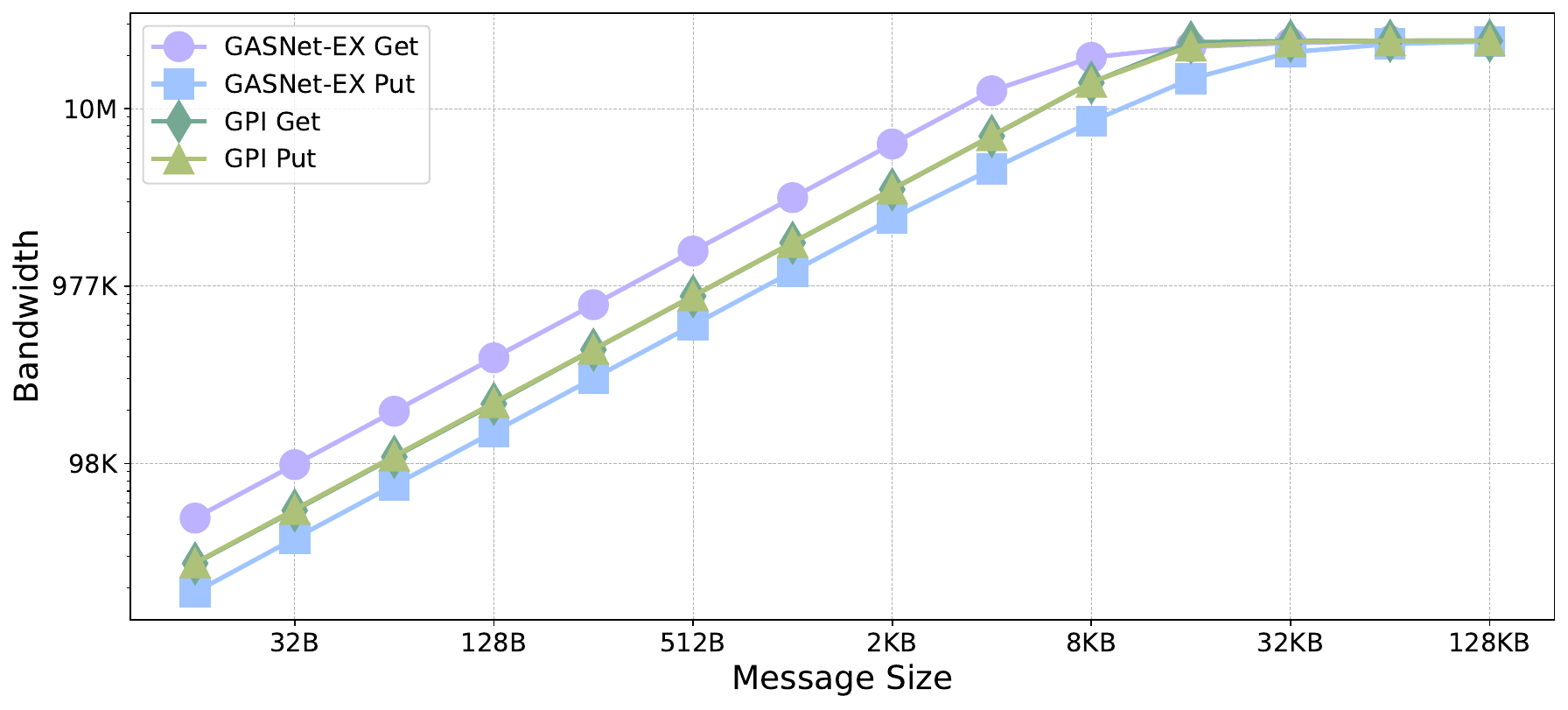}
    \vspace{-6 mm}
    \caption{Bandwidth comparison of two DiOMP implementations (GASNet-EX and GPI-2) over NDR InfiniBand.}
    \label{fig:gpi}
\end{figure}

To evaluate the efficiency of point-to-point communication, we conducted a micro-benchmark, focusing on latency and bandwidth performance across different communication modules. We compared the point-to-point performance of DiOMP RMA with MPI RMA. The results, shown in~\autoref{fig:lat} and~\autoref{fig:bw}, reveal that DiOMP outperforms MPI in nearly all scenarios, exhibiting significant advantages in both latency and bandwidth. The only exception is the DiOMP Put operation on Platform A\footnote{Platform A is an open platform, and the performance issue described has been documented in publicly available sources. However, disclosing this information here would violate ACM's double-blind review policy. If required, reviewers may contact the Program Chair through the review system, and we can provide relevant records upon request.}, which performs worse than MPI. Through communications with the respective platform administrators, vendors and developers, a hardware and driver-related issue has been identified. This issue has already been reported to HPE by the platform administrators. This issue is confirmed to be unrelated to DiOMP or the benchmark applications used in this study, and instead arise from external hardware and driver limitations. Importantly, these issues do not compromise the validity of the conclusions or contributions presented in this paper. In addition, we evaluated the performance of DiOMP's GPI-2 implementation in an InfiniBand environment. Figure~\ref{fig:gpi} shows the bandwidth comparison between GPI-2 and GASNet-EX. As we can see, GPI-2 outperforms GASNet-EX Put in certain scenarios. Overall, DiOMP exhibits significant performance advantages in point-to-point communication over MPI, highlighting its superior efficiency in RMA operations. DiOMP is capable of adapting to diverse hardware conditions while outperforming traditional MPI-based communication. 

\subsection{Micro-benchmark - Collective}

Then, to evaluate the performance of the two communication models in collective communication, we conducted latency tests for DiOMP and MPI on Broadcast and AllReduce operations across different data sizes. On platform A, we used 16 nodes, each equipped with 4 NVIDIA A100 GPUs, for a total of 64 GPUs. On platform B, we used 8 nodes, each equipped with 4 AMD MI250X GPUs, for a total of 32 GPUs (64 GCDs). On platform C, we utilized 4 nodes, each equipped with one NVIDIA Hopper GPU, for a total of 16 GPUs. 

\begin{figure}[htbp]
    \centering
    \subfigure[Broadcast]{
        \includegraphics[height=1.4in,width=.5\textwidth]{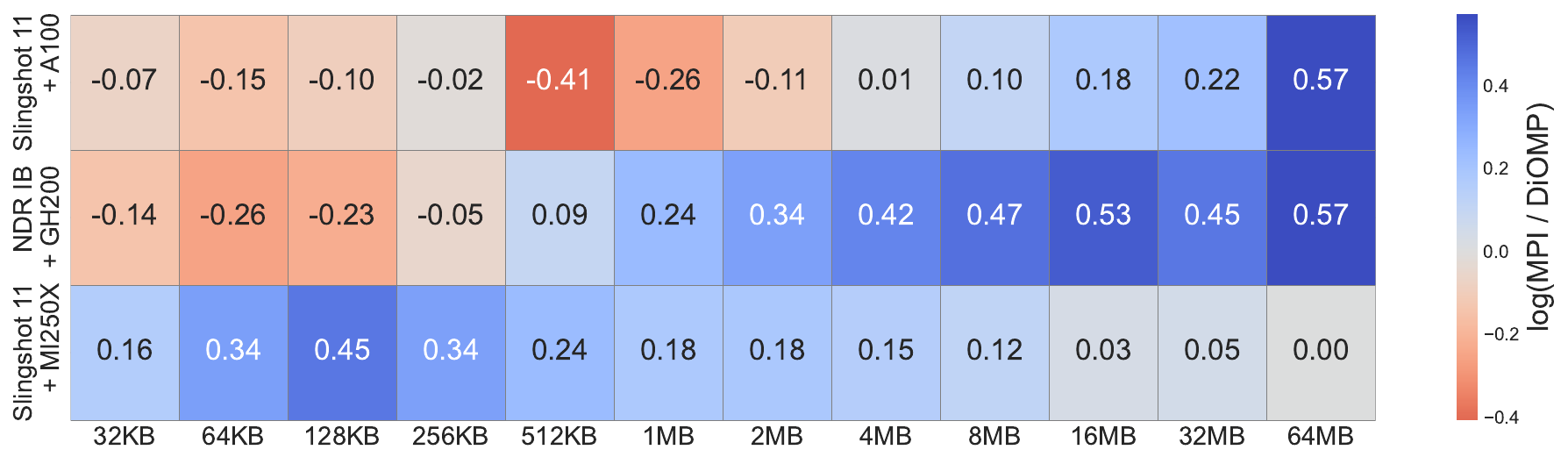}
    }
    \subfigure[ReduceAll(Sum)]{
        \includegraphics[height=1.4in,width=.5\textwidth]{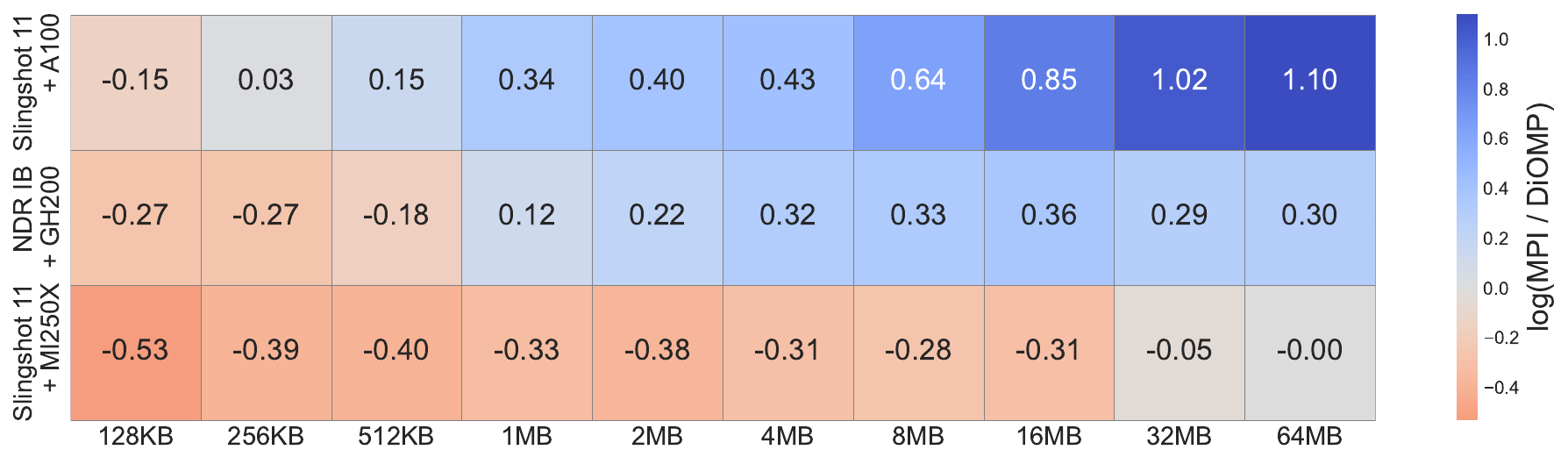}
    }
    \vspace{-5 mm}
    \caption{Logarithmic performance ratio ($log_{10}$) of MPI vs. DiOMP collective latency. Positive values (cool colors) indicate DiOMP is faster; negative values (warm colors) indicate MPI is faster.}
    \label{fig:coll}
\end{figure}

During the experiments, we first tested the Broadcast operation, 
we evaluated a range of data sizes from small (128 KB) to large (64 MB) to observe how latency varies with data size. Subsequently, we tested the AllReduce operation, in which data from all nodes is aggregated at the root node through addition. To ensure the stability of the results, each operation for a given data size was repeated 100 times, with the average latency reported as the final result. Additionally, to eliminate the impact of cold starts, multiple warm-up runs were conducted prior to the actual measurements. Figure~\ref{fig:coll} presents the relative collective communication time between MPI and DiOMP, expressed as $log_{10}\frac{MPI}{DiOMP}$. Warmer colors denote better MPI performance, while cooler colors indicate DiOMP is faster. When the amount of data is relatively small, DiOMP incurs higher latency compared to traditional MPI due to the overhead of initializing OMPCCL. On NCCL-based platforms A and C, DiOMP consistently demonstrates lower latency for large message sizes. On the RCCL-based platform B, DiOMP shows a noticeable advantage in broadcast operations for medium-sized messages. Although DiOMP achieves performance comparable to MPI for large message sizes, the performance gap between RCCL and MPI remains more pronounced than that between NCCL and MPI, suggesting that RCCL still has room for further optimization.

\begin{figure}[htbp]
    \vspace{-2 mm}
    \centering
    \subfigure[Slingshot 11 + A100]{
        \includegraphics[height=1.4in,width=.225\textwidth]{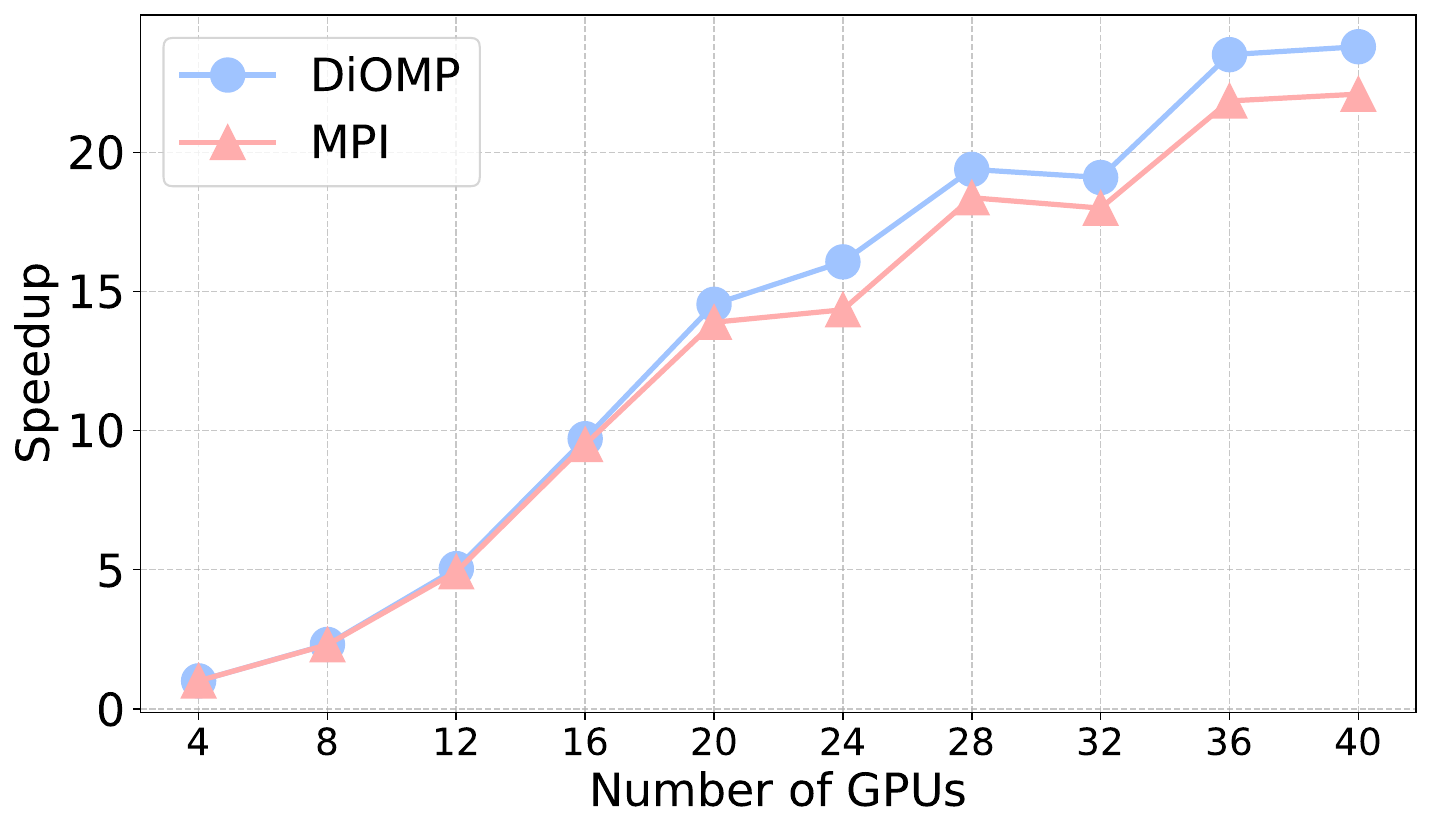}
    }
    \subfigure[Slingshot 11 + MI250X]{
        \includegraphics[height=1.4in,width=.225\textwidth]{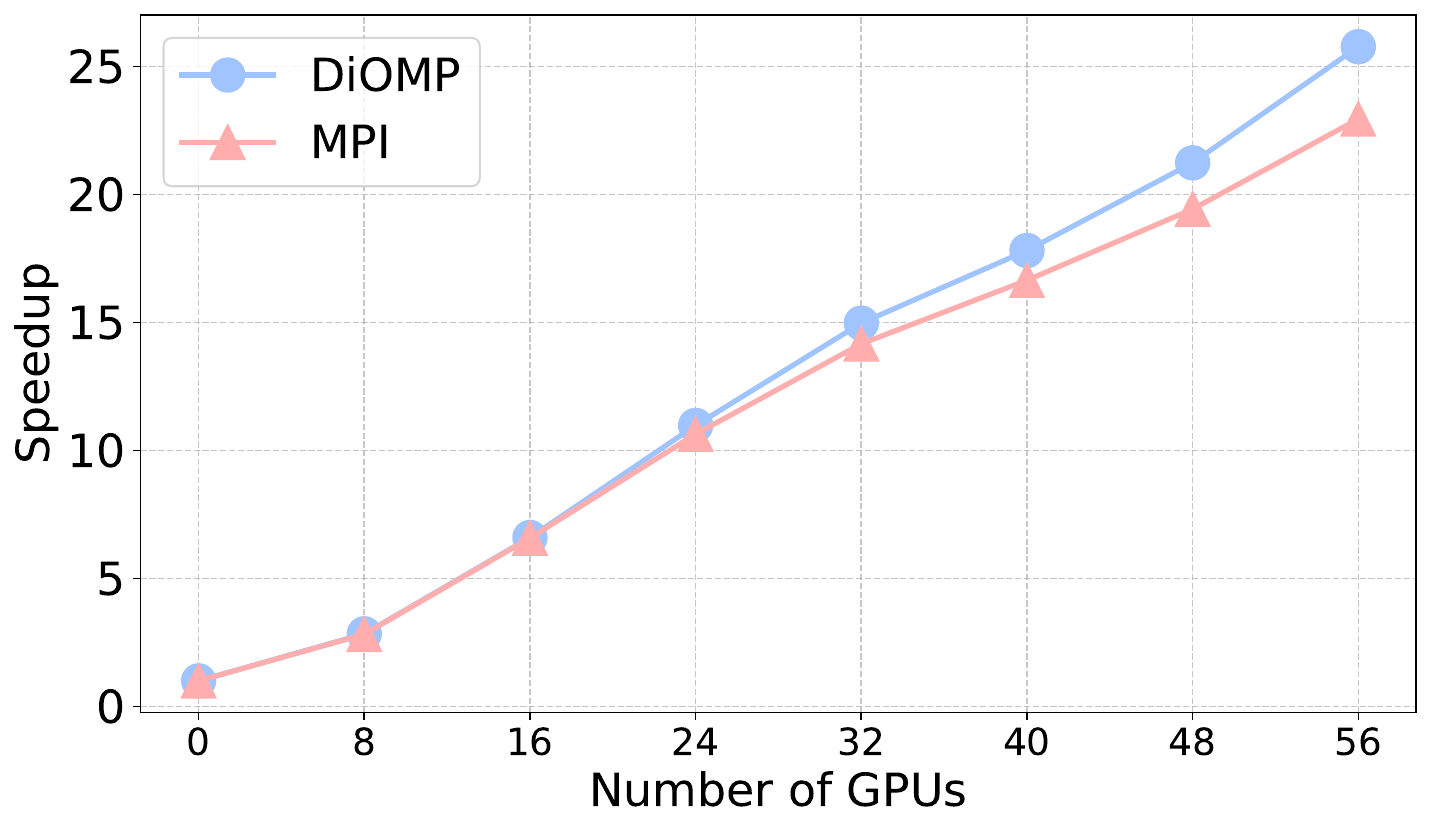}
    }
    \vspace{-5 mm}
    \caption{Matrix multiplication speedup on platform A and B with DiOMP and MPI+OpenMP. Higher is better.}
    \label{fig:res_mm}
\end{figure}

\subsection{Matrix Multiplication}
We subsequently tested the ring exchange communication pattern using an application that implements the Cannon algorithm to compute the square matrix product \( C = A \times B \). In the experiment, all three versions of the application utilized an additional block stripe for matrix \( B \) to enable overlap of computation and communication. Specifically, we set the number of processes(GPUs) as \( P \), the matrix size as \( N \), and the block stripe width as \( Ns = N / P \). During execution, each process (rank) completed \( P \) computations, with each computation involving a workload of \( N \cdot Ns \cdot Ns \). 

\autoref{fig:res_mm} presents the strong scaling results for multiplying two 
30240×30240 matrices on 4 to 40 NVIDIA A100 GPUs and 8 to 64 AMD MI250X GPUs. The baseline performance was measured using all GPUs within a single node—4 for A100 and 8 for MI250X—and speedups were computed relative to this baseline. During computation, communication latency was effectively masked, and the per-rank communication volume decreased as the number of GPUs increased. Consequently, the application exhibited superlinear scaling.





\subsection{Minimod}

Minimod~\cite{meng2020minimod} is a proxy application designed to simulate wave propagation in subsurface models by solving the finite-difference discretized wave equation using high-order stencil computation. In this study, we focus on the acoustic isotropic kernel. To adapt this kernel to distributed environments, we further developed a DiOMP-Offloading version. This implementation builds upon Minimod’s multi-GPU OpenMP Target version, with significant improvements to the halo exchange mechanism. Specifically, we replaced intra-node GPU-to-GPU communication with inter-node data transfers. This was achieved using DiOMP RMA, enabling seamless adaptation to distributed heterogeneous systems with minimal modifications to the original codebase. 
Listing~\ref{lst:minimod_diomp} and Listing~\ref{lst:minimod_mpi} present the pseudocode for the Halo Exchange portion of the Minimod program implemented with DiOMP and MPI+OpenMP, respectively. Analyzing the code reveals that compared to the MPI implementation, DiOMP significantly reduces programming complexity, requiring approximately half the lines of code to achieve equivalent data transfers. This highlights the simplicity and developer-friendliness of DiOMP, making it particularly suitable for HPC applications with complex data communication requirements.

\begin{listing}[htbp]
\begin{minted}[
  linenos=true,
  numbersep=5pt,
  xleftmargin=1.5em,
  breaklines=true,frame=single,
  fontsize=\footnotesize
]{c}
for (int r = 0; r < nranks; ++r) {
    llint gxmin, gxmax;
    RANK_XMIN_XMAX(r,gxmin,gxmax);
    if(rank == r) {
        if(rank != 0)
            ompx_put(...,D2D);
        if(rank != nranks - 1)
            ompx_put(...,D2D);
}}
ompx_fence();
\end{minted}
\vspace{-3 mm}
\caption{Halo Exchange Code of Minimod with DiOMP}
\label{lst:minimod_diomp}
\end{listing}

\begin{listing}[htbp]
\begin{minted}[
  linenos=true,
  numbersep=5pt,
  xleftmargin=1.5em,
  breaklines=true,frame=single,
  fontsize=\footnotesize
]{c}
MPI_Request requests[4];
int req_cnts = 0;
for (int r=0; r<nranks; r++) {
    RANK_XMIN_XMAX(r,gxmin,gxmax);
    if (rank == r) {
        if (r != 0) {
            #pragma omp target data use_device_ptr(v)
            MPI_Isend(..., &requests[req_cnts++]);
        } if (r != nranks-1) {
            #pragma omp target data use_device_ptr(v)
            MPI_Isend(..., &requests[req_cnts++]);
        }
    } if (rank == r-1) {
        #pragma omp target data use_device_ptr(v)
        MPI_Irecv(..., &requests[req_cnts++]);
    }
    if (rank == r+1) {
        #pragma omp target data use_device_ptr(v)
        MPI_Irecv(..., &requests[req_cnts++]);
}}
MPI_Waitall(req_cnts, requests, MPI_STATUSES_IGNORE);
\end{minted}
\vspace{-3 mm}
\caption{Halo Exchange Code of Minimod with MPI}
\label{lst:minimod_mpi}
\end{listing}

\begin{figure}[htbp]
    \vspace{-3 mm}
    \centering
    \subfigure[Slingshot 11 + A100]{
        \includegraphics[height=1.4in,width=.225\textwidth]{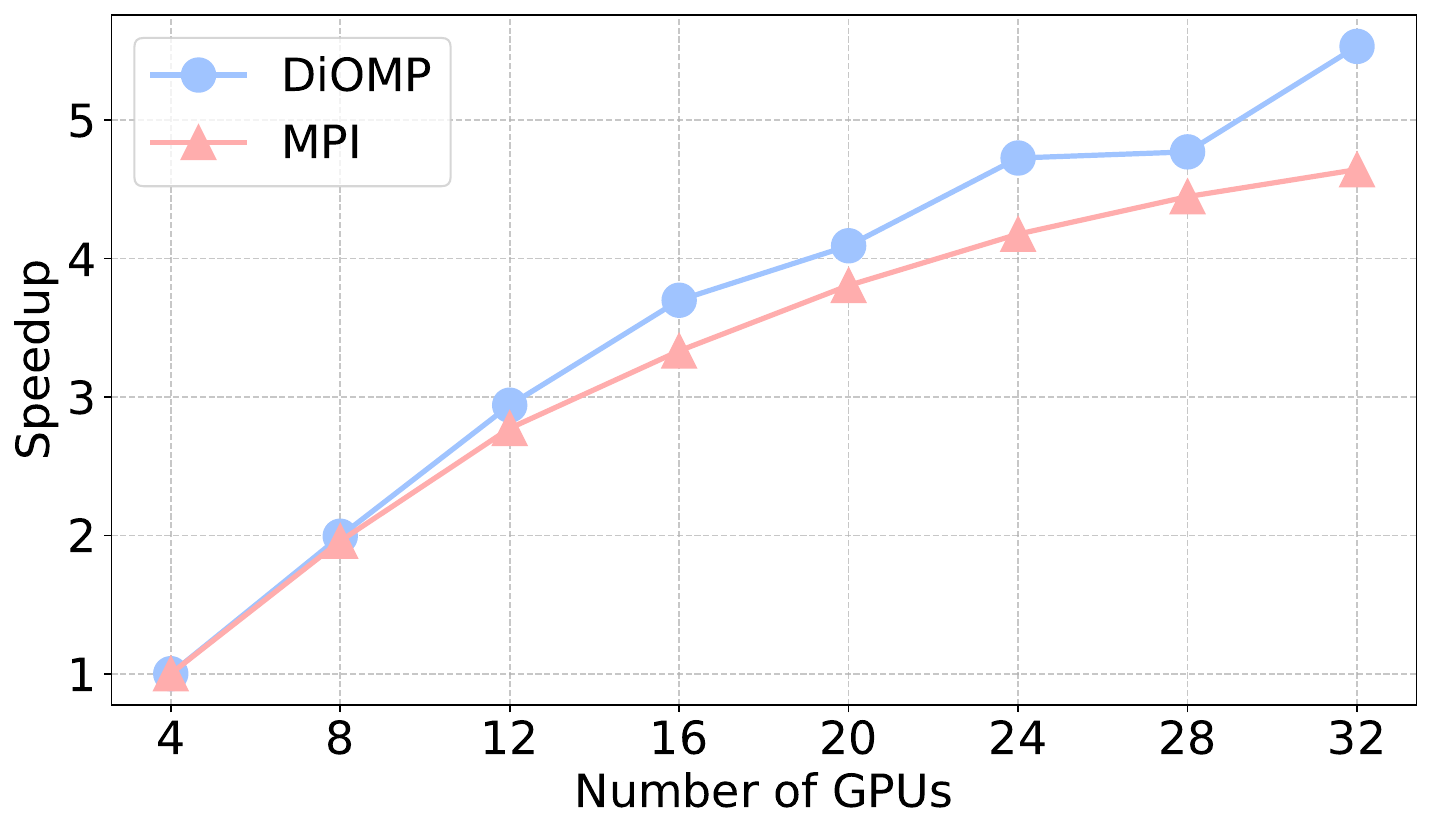}
    }
    \subfigure[Slingshot 11 + MI250X]{
        \includegraphics[height=1.4in,width=.225\textwidth]{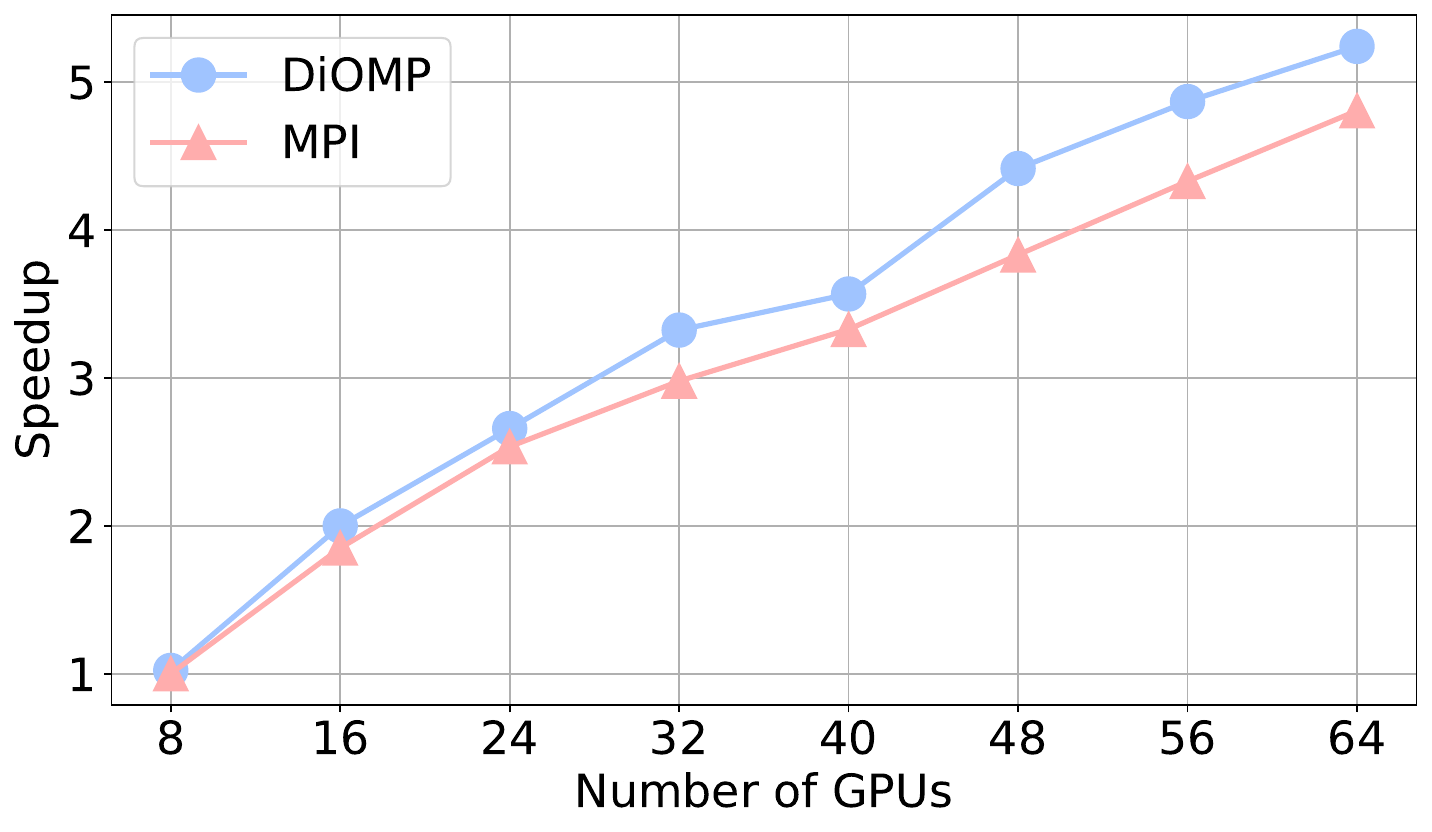}
    }
    \vspace{-4 mm}
    \caption{Minimod speedup comparison of DiOMP and MPI on using HPE Slingshot 11 and InfiniBand. Higher is better.}
    \label{fig:res_minimod}
\end{figure}

\vspace{-2 mm}
As shown in ~\autoref{fig:res_minimod}, for a grid size of $1200^3$ and 1000 steps, the speedup trends of the Minimod program across different platforms with varying GPU counts are demonstrated. Benefiting from DiOMP’s optimized intra-node communication mechanisms, the DiOMP implementation demonstrates superior performance over MPI in single-node, multi-device environments. Accordingly, we adopt MPI’s single-node performance as the baseline for all speedup evaluations. Experimental results across various node counts and platforms consistently show that DiOMP outperforms the MPI implementation of Minimod with significantly reduced code complexity and programming effort. These results highlight DiOMP’s ability to simultaneously deliver high performance and improved programmability, making it a compelling alternative to traditional MPI-based approaches.

\section{Conclusion and Future Work}
\label{sec:conclusion}
In this work, we introduced DiOMP-Offloading, a unified programming framework that efficiently supports OpenMP execution in heterogeneous, multi-node HPC environments. By integrating PGAS-style data distribution, OpenMP target offloading, and our novel OMPCCL communication abstraction, the framework achieves both high programmability and superior performance. Experimental results across various platforms confirm that DiOMP-Offloading transparently handles inter-device communication, significantly reducing manual effort and improving performance in representative applications such as matrix multiplication and Minimod.

While DiOMP-Offloading has demonstrated clear advantages in scalability and communication efficiency, one promising direction for future work is the integration of task-level parallelism in OpenMP. As modern applications increasingly rely on dynamic and irregular execution patterns, extending DiOMP-Offloading to support task-based parallelism within the PGAS model will further enhance its flexibility and applicability. Additionally, ongoing efforts aim to strengthen compiler-level integration with LLVM, enabling automated optimizations for remote data access and memory management, and thereby reducing programming complexity.


\bibliographystyle{ACM-Reference-Format}
\bibliography{offloading}


\end{document}